# Structure and Control of Charge Density Waves in Two-Dimensional 1T-TaS$_2$

A. W. Tsen[1], R. Hovden[2], D. Z. Wang[3], Y. D. Kim[4], J. Okamoto[5], K. A. Spoth[2], Y. Liu[6], W. J. Lu[6], Y. P. Sun[6,7,8], J. Hone[4], L. F. Kourkoutis[2,9], P. Kim[1,10*], and A. N. Pasupathy[1*]

[1]Department of Physics, Columbia University, New York, New York 10027, USA

[2]School of Applied and Engineering Physics, Cornell University, Ithaca, New York 14850, USA

[3]Department of Applied Physics and Applied Mathematics, Columbia University, New York, New York 10027, USA

[4]Department of Mechanical Engineering, Columbia University, New York, New York 10027, USA

[5]Department of Physics, University of Hamburg, D-20355 Hamburg, Germany

[6]Key Laboratory of Materials Physics, Institute of Solid State Physics, Chinese Academy of Sciences, Hefei 230031, People's Republic of China

[7]High Magnetic Field Laboratory, Chinese Academy of Sciences, Hefei 230031, People's Republic of China

[8]Collaborative Innovation Centre of Advanced Microstructures, Nanjing University, Nanjing 210093, People's Republic of China

[9]Kavli Institute at Cornell for Nanoscale Science, Ithaca, New York 14853, USA

[10]Department of Physics, Harvard University, Cambridge, Massachusetts 02138, USA

*Correspondence to: pkim@physics.harvard.edu, pasupathy@phys.columbia.edu

**The layered transition metal dichalcogenides host a rich collection of charge density wave (CDW) phases in which both the conduction electrons and the atomic structure display translational symmetry breaking. Manipulating these complex states by purely electronic methods has been a long-sought scientific and technological goal. Here, we show how this can be achieved in 1T-TaS$_2$ in the two-dimensional (2D) limit. We first demonstrate that the intrinsic properties of atomically-thin flakes are preserved by encapsulation with hexagonal boron nitride in inert atmosphere. We use this facile**

**assembly method together with TEM and transport measurements to probe the nature of the 2D state and show that its conductance is dominated by discommensurations. The discommensuration structure can be precisely tuned in few-layer samples by an in-plane electric current, allowing continuous electrical control over the discommensuration-melting transition in 2D.**

1T-TaS$_2$ is a layered material that exhibits a number of unique structural and electronic phases. At low temperature and ambient pressure, the ground state is a commensurate (C) charge density wave (CDW). Upon heating, it undergoes a sequence of first order phase transitions to a nearly commensurate (NC) CDW at 225K, to an incommensurate (IC) CDW at 355K, and finally to a metallic phase at 545K. Each transition involves both conduction electron and lattice degrees of freedom—large changes in electronic transport properties occur, concomitant with structural changes to the crystal. By either chemical doping or applying high pressures, it is possible to suppress the CDWs and induce superconductivity[1-3]. For device applications, it is desirable to control these phases by electrical means, but this capability is difficult to achieve in bulk crystals due to the high conduction electron density. Recent efforts to produce thin samples by mechanical exfoliation provide a new avenue for manipulating the CDWs in 1T-TaS$_2$[4-7]. These studies have demonstrated the suppression of CDW phase transitions using polar electrolytes as well as resistive switching between the different phases. As the material approaches the two-dimensional (2D) limit, however, significant changes have been observed in the transport properties[4,5]. Yet, the microscopic nature of the 2D state remains unclear. In this work, we use transmission electron microscopy (TEM) together with transport measurements to develop a systematic understanding of the CDW phases and phase transitions in ultrathin 1T-TaS$_2$. We find that charge ordering disappears in flakes with few atomic layers due to surface oxidation. When samples are instead environmentally protected, the CDWs persist and their transitions can be carefully tuned by electric currents.

Both the atomic and CDW structure of 1T-TaS$_2$ can be visualized in reciprocal space by TEM electron diffraction[8,9]. In Figure 1a, we show diffraction images taken from a bulk-like, 50nm thick crystal at low and room temperature (C phase–blue panel, NC phase–red panel). The bright peaks (connected by dashed lines) correspond to Bragg scattering from a triangular lattice of Ta atoms with lattice constant $a$ = 3.36Å. Additional weaker diffraction peaks appear from the periodic atomic displacements of the CDW. In the low-temperature C phase, Ta atoms displace to make Star-of-David clusters (blue inset, Fig. 1b). The outer twelve atoms within each star displace slightly inwards towards the atom at the center, giving rise to a commensurate superstructure with wavelength $\lambda_C = \sqrt{13}a$ that is rotated $\phi_C$ = 13.9º with respect to the atomic lattice. The NC phase at room temperature also consists of such 13-atom distortions. Scanning tunneling microscope (STM) measurements have revealed, however, that such ordering is only preserved in quasi-hexagonal domains consisting of tens of stars[10,11], with domain periodicity 60Å to 90Å depending on temperature[12,13]. The domains are separated by a discommensuration network forming a Kagome lattice, inside of which the Ta displacements are substantially reduced[14]. A schematic of this structure is shown in the red inset of Figure 1b.

When ultrathin 1T-TaS$_2$ crystals (c.a. < 5nm thickness) are exfoliated in an ambient air environment, the CDW structure is not observed by the TEM electron diffraction. In the left panel of Figure 1c, we show a room temperature electron diffraction pattern taken on a few-layer flake. The presence of Bragg peaks without CDW scattering suggests that the 1T-TaS$_2$ layers are in a phase that is not observed in bulk crystals at this temperature. High-resolution electron microscopy and energy dispersive spectroscopy on fully suspended samples reveal a strong presence of oxidation as well as an amorphous layer on the surface (see Supplementary Information). The amorphous oxide (~ 2nm thickness) can be clearly seen atop both surfaces of the 1T-TaS$_2$ layers in cross-section (Fig. 1c, right panel). It is possible that oxidation leads to strong surface pinning, which destroys charge ordering in ultrathin samples. Recent resistivity measurements on exfoliated 1T-TaS$_2$ crystals have also reported the disappearance of CDWs in sufficiently thin flakes[5]. It is not clear, however, whether these are intrinsic effects related to dimensionality or extrinsic consequences of oxidation.

In order to prevent surface oxidation, we exfoliated 1T-TaS$_2$ crystals within a nitrogen-filled glove box with under 2ppm oxygen concentration. The flakes were protected by a capping layer of thin hexagonal boron nitride (hBN) before transfer out into the ambient environment (see Methods). TEM diffraction performed on these protected samples reveals that CDW formation persists down to the lowest thicknesses measured (2nm), as we discuss in detail in Figure 4. This indicates that the absence of charge order in ultrathin, uncovered flakes is most likely caused by the effects of oxidation. The study and utilization of CDWs in 2D 1T-TaS$_2$ thus requires careful sample preparation in inert atmosphere.

The different structural phases of 1T-TaS$_2$ exhibit distinct electronic transport properties that may be exploited for device applications. In the main panel of Figure 1b, we show temperature-dependent resistivity of a bulk crystal measured across the NC-C phase transition. Resistivity abruptly increases (decreases) by over an order of magnitude upon entering the C (NC) phase. The hysteresis loop between cooling and warming defines the temperature region of metastability between the two phases and can be understood by a free energy picture (Fig. 1d). In a first-order transition, an activation barrier separates the stable energy minima corresponding to the NC and C states. With cooling from the NC phase, both the C state energy and the height of the barrier decrease with respect to the NC energy. When the C state has lower energy, the NC phase becomes metastable, but the system only transitions into the C phase when the activation barrier becomes comparable to the thermal energy. The situation is reversed when warming from the C phase. In oxidation-free 2D samples, this electronic transition is qualitatively unchanged.

Figure 2a shows an example of hBN-encapsulated 1T-TaS$_2$ flakes before (top) and after device fabrication (bottom). In order to make electrical contact to the covered samples, we used a technique of edge metallization developed for graphene/hBN heterostructures (see Methods)[15]. A side-view device schematic is shown in the inset of the bottom panel. In the main panel of Figure 2b ($I = 0$, black curve), we plot resistance as a function of temperature for a 4nm thick sample measured across the NC-C phase transition. The behavior is similar to that of the bulk crystal (Fig. 1b); however, the hysteretic region between cooling and warming is substantially widened, indicating that one or both of the CDW phases become more metastable.

Metastable phases of a CDW system are generally more susceptible to electronic perturbations, since CDWs directly couple to electric field[16,6,7]. In our device, we observe that continuous current flow stabilizes the NC phase at low temperatures. In Figure 2b (main panel), we show AC resistance with temperature while also applying a continuous, in-plane DC current, starting at room temperature. As the DC current $I$ is increased, the final resistance at low temperature is monotonically lowered. Concomitant with this trend, the resistance jump resulting from the NC-C phase transition also decreases with increasing $I$. In the inset, we have plotted the ratio of the resistance difference between cooling and warming, $\Delta R$, to resistance $R$ in the more conducting state at $T = 180$K, the temperature in the middle of the hysteresis region, as a function of the DC current level. For $I = 35\mu$A (blue curve in main panel), the NC-C phase transition is completely absent. This indicates that C phase formation in the current driven sample is very different compared to the zero-current, equilibrium condition. Current flow hinders the formation of the C phase and maintains the sample in the more conductive NC state at low temperature. We exclude Joule heating of the sample as a possible explanation by slowly turning off the current at low temperature and verifying that the resistance does not change. Our observation suggests that it is possible to stabilize the NC phase in a temperature region where it is not thermodynamically stable, and thus substantially increase the lifetime of the metastable state.

While continuous current flow can suppress the NC-C phase transition, we find that the opposite phenomenon is also possible, i.e. we can drive a transition towards the thermodynamically stable state, if we apply a transient in-plane current instead. Figure 2c shows the current induced phase transitions in the same device (4nm thickness). Here, we start in the NC phase at room temperature and cool the sample down to 150K without current flow. At this temperature, while the sample remains in the NC state, the NC phase is now metastable, and the C phase is the thermodynamically stable state. As we increase the voltage across the device (top panel, dark green curve), the measured current through the device decreases in abrupt steps (marked by red arrows) when it reaches a critical current $I_c \sim 30\mu A$ (marked by red dashed line). Upon sweeping the bias current back to zero (light green curve), the device remains in a more insulating state. Warming up the sample after this point produces a temperature curve similar to the C phase, and a transition to the NC phase is observed. This demonstrates that a bias current applied to the sample can be used to drive the metastable NC phase towards the thermodynamically preferred C state. We have used the dashed green arrow in Figure 2b to mark the direction of this current-induced NC to C phase transition and a free energy schematic of this process is shown in the inset of the upper panel of Figure 2c. We should emphasize that this current driven NC to C transition occurs only for $I > I_c$ applied transiently after cooling under equilibrium conditions. Cooling with continuous current flow starting at room temperature would not result in the induced transition, but stabilizes the NC phase down to lowest temperatures measured.

Similarly, the metastable C state can also be driven towards the NC phase with current. Here, we start in the C phase at 50K and warm up to 200K. The sample remains in the C phase, but now the NC phase is the thermodynamic ground state. As shown in the lower panel of Figure 2c, sweeping the voltage in this case results in a sharp increase in current and drives the sample towards the more conducting NC state. We have used the dashed orange arrow in Figure 2b and the free energy picture in the inset of the lower panel of Figure 2c to represent this opposite C to NC transition. Interestingly, both induced transitions occur when the current reaches about $I_c \sim$ 30µA, indicating that indeed current flow rather than electric field is the underlying mechanism that drives the transition. We have repeated this measurement at various temperatures and initial conditions. In all cases, whenever the initial system is metastable, reaching a current threshold of 30µA to 40µA drives the system to the thermodynamically stable state, regardless of device resistance. In contrast, we observe no induced transition up to 45µA at 260K, where a metastable phase ceases to exist.

Taken together, the results of Figure 2 demonstrate that it is possible to electrically control the NC-C transition in 2D 1T-TaS$_2$, where the temperature region of metastability is significantly enhanced. A more detailed study of this phase transition in 2D samples, however, can provide a better understanding of our experimental observations. The key structural difference between the two CDW phases is the presence of the discommensuration network in the NC phase (Fig. 1b, red inset). The NC-C transition can then be interpreted as a discommensuration-melting transition, which can be significantly affected by dimensionality[17,18]. The discommensurations have a striking effect on the electronic transport properties in 1T-TaS$_2$. The NC phase is an order of magnitude more conductive than the C phase. If we assume that the interior of each commensurate domain has similar transport properties as the C phase, this then implies the discommensuration regions in the NC phase are at least ten times more conductive than the domain interior[3]. Such a view is supported by the fact that the atomic structure within the discommensurations is close to the high-temperature metallic phase[14]. With this interpretation, we can use transport measurements to better understand the role of dimensionality on the discommensuration-melting transition.

As the number of 1T-TaS$_2$ layers decreases, the resistivity change corresponding to the NC-C phase transition evolves in a continuous manner down to 2nm thickness in environmentally protected samples. Figure 3a shows resistivity as a function of temperature for four, hBN-covered 1T-TaS$_2$ flakes, all measured using a 1K/min sweep rate. Their thicknesses are 2, 4, 6, and 8nm as determined using an atomic force microscope. For comparison, we show data from an unprotected, 20nm thick flake, which exhibits characteristics similar to the bulk crystal, indicating that the effects of oxidation are less pronounced in thicker samples. The temperature hysteresis associated with the phase transition between cooling and warming is substantially increased in thinner samples, consistent with our earlier observations of the device in Figure 2a. The progressive widening of the hysteresis loop continues down to the 4nm thick device, below which there is no longer a detectable transition. A guide to the eye for the expansion of this metastable region is shown by the colors in Figure 3a. In the upper panel of Figure 3b, we plot $\Delta T = T_{c,warm} - T_{c,cool}$ as a function of flake thickness, where $T_{c,warm}$ and $T_{c,cool}$ are the experimentally observed NC to C or C to NC transition temperature during the warming or cooling process, respectively. Here, $T_c$ is determined by the temperature at which the first derivative peaks in the temperature sweep. $\Delta T$ is 60K for the 20nm flake, slightly larger than that for the bulk crystal (40K), and grows to 120K for the 4nm device. The average temperature $T_{c,avg}$ = $(T_{c,warm}+T_{c,cool})/2$ of the transition, however, does not change substantially with thickness and remains between 180K and 190K. This then implies that lower dimensionality does not stabilize either the NC or C phase. Instead, the NC (C) phase becomes increasingly metastable during cooling (warming) for thinner samples, indicating that the size of the energy barrier separating the NC and C phases increases (Fig. 1d).

While $\Delta T$ decreases with decreasing thickness of the sample, the resistivity difference, $\Delta\rho$, between cooling and warming decreases with decreasing sample thickness as well. In the bottom panel of Figure 3b, we plot the ratio of $\Delta\rho$ to $\rho$ in the more conducting state at $T = 180$K as a function of flake thickness. For the 20nm device, resistivity changes by close to an order of magnitude, slightly less than in the bulk crystal shown in Figure 1b. The change is smaller for thinner devices and disappears completely for the 2nm device, suggesting no phase transition has occurred. This indicates that more conducting NC discommensurations persist at low temperatures for decreasing sample thickness, consistent with the larger energy barriers required to remove them. Also, the resistivity jump becomes less abrupt in thinner samples, which is a reflection that the phase transition has slowed, as larger energy barriers generally act also to impede the kinetics of a phase transition. A simple circuit model presented in Figure 3c allows us to connect the measured resistance jump in the NC-C transition, $\Delta R$, with the estimated density of discommensurations $1/d$ left in the low temperature phase. We assume that the device resistance at low temperature is dominated by conduction through a random network of discommensuration channels (shown as white lines), which is generally sensitive to the particular microstructure formed. However, for device sizes much larger than $d$, we find the resistance with discommensuration channels would be $R \sim \rho_{DC} d$ where $\rho_{DC}$ is the resistivity per unit length of each discommensuration channel. Similarly, in the high temperature NC phase with a well-defined discommensuration network, we have $R_{NC} \sim \rho_{DC} D_{NC}$, where we assume $D_{NC} \sim 80\text{Å}^{12,13}$. From this, we can use the resistivity change in Figure 3b to determine $d$: $\frac{\Delta R}{R_{NC}} \sim \frac{d}{D_{NC}} - 1$. On the right axis, we have plotted $d$ extracted for the different sample thicknesses. For the 2nm sample, $d \sim D_{NC}$, while it grows to $d \sim 700\text{Å}$ for the 20nm sample.

We can further substantiate the microscopic picture presented above by providing atomic structural analysis based on TEM. As before, the CDW structure is maintained by environmentally controlled hBN encapsulation. In Figure 4a, we show diffraction images taken from two 1T-TaS$_2$ flakes of different thicknesses (12nm and 2nm). To highlight their temperature dependence, we have overlaid the diffraction patterns for each flake at 295K (red peaks) and 100K (blue peaks), our lowest achievable temperature. Ta Bragg peaks are again connected by a dashed triangle. Multiple scattering from hBN creates additional discernable peaks. The CDW peaks have been circled for easy identification. While the peaks circled in gray appear qualitatively similar for both flakes, only the thicker flake displays additional peaks (circled in blue) at 100K, indicating that it makes the transition to the C phase (compare with blue panel in Figure 1a), while the thinner flake remains in the NC phase. This is consistent with our transport data as larger energy barriers in thinner samples require lower temperatures to realize the C phase.

The movement of the gray-circled peaks with cooling (denoted by arrows, Fig. 4a) can be understood more quantitatively with reference to the zoom-in schematic shown in Figure 4b (upper right). The position of this CDW peak is related to the periodicity $D_{NC}$ of the NC domains

(upper left) by a simple geometric expression[13]: $D_{NC} = \frac{a}{\sqrt{\left(\frac{2\pi\Delta\phi}{360°}\right)^2 + \left(\frac{\Delta\lambda}{\lambda_C}\right)^2}}$, where $\Delta\phi$ is the difference in degrees between $\phi$ and $\phi_C = 13.9°$ and $\Delta\lambda$ is the difference between the apparent wavelength averaged over many domains and $\lambda_C = \sqrt{13}a$. Thus, as the domain size grows, the NC peaks move closer to the C phase positions. We have explicitly measured the position and angle of the CDW wavevectors for these two samples at several different temperatures during cooling in order to determine the domain period $D_{NC}$ using the expression above. The results are plotted in the lower panel of Figure 4b. For comparison, we also reproduce STM results obtained by Thomson *et al.* on the surface of a bulk crystal[13]. For bulk samples, $D_{NC}$ grows steadily from 60Å to 90Å upon cooling from 340K to 215K and then jumps to an arbitrarily large value upon transition into the C phase at ~ 180K. At the same time, the width of the discommensuration regions remains relatively constant (~ 22Å) in all of the NC phase[12]. As with our transport results, we find that reducing sample thickness suppresses the NC to C phase transition to lower temperatures during cooling and slows the CDW domain growth rate during the transition. For both of the thin flakes, the initial domain size at room temperature is similar to that that of the bulk crystal ($D_{NC}$ = 60Å to 70Å). $D_{NC}$ increases slightly upon cooling in the NC phase. For the 12nm flake, the C phase is formed between 100K to 150K, whereas the 2nm flake remains in the NC phase even at 100K. Its domain size here is much larger ($D_{NC}$ ~ 500Å), however, indicating that the phase transition has begun to take place. This is in clear contrast to bulk samples where the transition is abrupt.

Our transport and TEM measurements both indicate that reduced dimensionality increases the energy barrier separating the NC and C CDW phases, and thus widens the metastable region of the phase transition. The transition into the C phase involves melting or removal of the NC discommensuration network. Microscopically, energy barriers to discommensuration motion have been attributed to the presence of defects or impurities in the material which act to pin them locally[19]. Even in nominally pure CDW samples, clusters of localized defects have been observed by STM[20,21], where the distance between defects is on the order of ~ 10nm. In bulk 1T-TaS$_2$, the interlayer stacking of NC domains make the discommensuration walls extended planar objects[14,22], which are generally more difficult to pin. In 2D, however, the discommensurations become lines, which may be more easily immobilized. We have constructed a model of discommensuration pinning for a 2D system of thickness $t$ (see Supplementary Information). We find that in the ultrathin limit where $t$ is smaller than the mean distance between impurities, the pinning energy for a discommensuration plane scales as $E_{pin} \sim t^{-2/3}$, corresponding to a crossover from collective weak pinning to strong individual pinning. These strong pinning centers stabilize the NC discommensuration network at low temperatures during cooling, and will also hinder the nucleation and growth of discommensurations when warming from the C phase. This increases the temperature region of metastability for both CDW phases in accordance with our experimental observation.

It is most likely that the larger pinning energies in 2D allow us to effectively control the NC-C phase transition in few-layer samples. By flowing sufficient current transiently through a sample in a metastable CDW state, we may depin the discommensurations and assist the formation of the thermodynamic ground state (Fig. 2c). On the other hand, if we continuously flow current starting at room temperature, we may help to stabilize the conducting NC discommensuration network even upon cooling to low temperatures (Fig. 2b). Although a spatially-resolved study is still needed to fully understand these effects, our results have both elucidated the nature of the 2D state in 1T-TaS$_2$ as well demonstrate clear electrical control over the NC-C phase transition in ultrathin samples, further establishing the material's relevance for device applications. We also expect our environmentally controlled techniques to be applicable for the study of other 2D transition-metal dichalcogenides which may be unstable under ambient conditions[23].

**Methods:**

*Synthesis of 1T-TaS$_2$*

High-quality single crystals of 1T-TaS$_2$ were grown by the chemical vapor transport (CVT) method with iodine as a transport agent. The high-purity Ta (3.5N), S (3.5N) and Se (3.5N) were mixed in chemical stoichiometry, and heated at 850°C for 4 days in an evacuated quartz tube. The harvested TaS$_2$ powders and iodine (density: 5 mg/cm$^3$) were then sealed in an another quartz tube and heated for two weeks in a two-zone furnace, in which the source zone and growth zone were fixed at 900°C and 800°C, respectively. The tubes were rapidly quenched in cold water to ensure retaining of the 1T phase.

*Device Assembly and Fabrication*

We exfoliated thin 1T-TaS$_2$ flakes onto SiO$_2$/Si wafers inside a N$_2$-filled glovebox containing below 2ppm O$_2$ concentration. Outside the glovebox we separately exfoliated single-crystal hBN flakes onto SiO$_2$/Si. Using a polydimethylsiloxane (PDMS) stamp covered with polypropylene carbonate (PPC), we "picked up" thin hBN (< 30nm thickness) via the method described in Wang *et al.*[15]. This sample was then moved inside the glove box. In order to prepare the 1T-TaS$_2$ for TEM study, we used the hBN to again pick up 1T-TaS$_2$ *in-situ* and then transfer the hBN/1T-TaS$_2$ stack onto a TEM chip with a SiN membrane. The chip was then moved outside of the glove box and cleaned in acetone.

In order to prepare 1T-TaS$_2$ for transport studies, we again exfoliated flakes on SiO$_2$/Si inside the glove box. hBN was used to cap 1T-TaS$_2$ in a manner similar to that described above.

To make electrical contact to the covered flakes, electron beam resist was used as an etch mask to pattern the device channel, leaving 1T-TaS$_2$ exposed at the edges. A subsequent lithography step was then used to pattern metal electrodes contacting the 1T-TaS$_2$ edge. We performed a second etch immediately prior to metal evaporation (1nm Cr/50nm Au) to expose a new edge with reduced oxidation. Alternatively, we have also "picked up" thin 1T-TaS$_2$ with hBN/graphene heterostructures and placed the entire stack on another hBN substrate. Here, few-layer graphene was used as electrodes to contact the top surface of 1T-TaS$_2$. We see no substantial difference in the temperature-dependent resistivity behavior between these two processes.

*Scanning / Transmission Electron Microscopy*

The conventional transmission electron microscopy experiments, including electron diffraction, were conducted on an FEI T12 Bio-Twin operating at 80keV. Temperatures as low as ~97K were accessible using a cryogenic specimen stage. A high vacuum environment and the microscope's fixed cryo-shields prevented accumulation of ice on the specimen when held at low temperatures. High-resolution scanning transmission electron microscopy images were collected on a FEI Tecnai F20 transmission electron microscope operating at 200keV with a probe forming semi-angle of roughly 9.6mrad and a high-angle annular dark field detector at a camera length of 150mm.

**Acknowledgements:**

The authors thank F. J. DiSalvo, R. E. Thorne, A. J. Millis, I. L. Aleiner, and B. L. Altshuler for useful discussions.

This material is based upon work supported by the NSF MRSEC program through Columbia in the Center for Precision Assembly of Superstratic and Superatomic Solids (DMR-1420634). PK acknowledges support from ARO (W911NF-14-1-0638).

This work was supported by the National Key Basic Research under Contract No. 2011CBA00111, the National Nature Science Foundation of China under Contract No. 11404342, the Joint Funds of the National Natural Science Foundation of China and the Chinese Academy of Sciences' Large-scale Scientific Facility (Grand No. U1232139).

The work at Cornell was supported by the David and Lucile Packard Foundation.

**Author contributions:**

A.W.T, P.K., and A.N.P. conceived and designed the experiment. Y.L. and W.J.L. synthesized the $1T$-$TaS_2$ crystals. A.W.T. fabricated the samples with assistance from D.Z.W. and Y.D.K. A.W.T. performed the transport measurements. A.W.T. and R.H. performed the TEM diffraction measurements with assistance from K.A.S. R.H. performed the STEM, EDX, and EELS measurements. J.O. provided the pinning energy analysis. A.W.T, P.K., and A.N.P. analyzed the data and wrote the paper.

**Competing financial interests:**

The authors declare no competing financial interests.

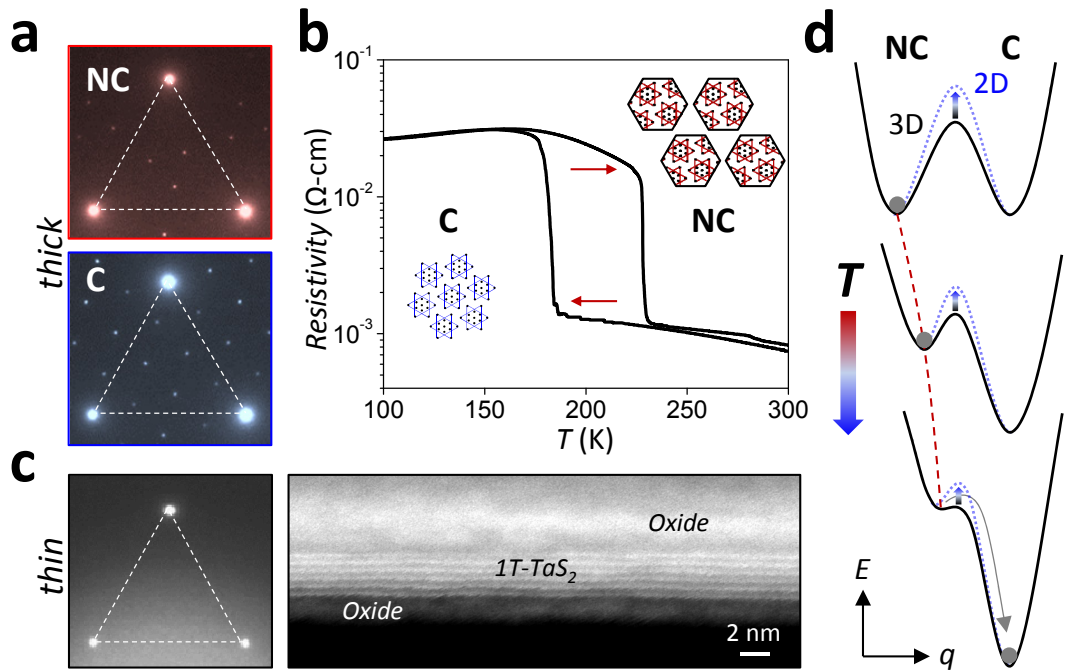

**Figure 1 | NC-C CDW phase transition in bulk 1T-TaS$_2$ and CDW suppression by oxidation in thin flakes. a.** TEM diffraction images of 50nm thick 1T-TaS$_2$ at 295K (red, NC phase) and 100K (blue, C phase). Weaker peaks are due to CDW distortion. **b.** Resistivity vs. temperature of bulk 1T-TaS$_2$ crystal around the first-order, NC-C transition. Insets show real space schematics of CDW structure. **c.** (Left) TEM diffraction of few-layer 1T-TaS$_2$ flake shows absence of CDW order. (Right) High-resolution, cross-section electron microscopy image reveals presence of amorphous oxide. **d.** Free energy schematic of CDW evolution with temperature. Vertical and horizontal axis represent free energy ($E$) and reaction coordinate ($q$), respectively. NC domains grow slowly upon cooling until abrupt transition into the C phase. Energy barrier increases in 2D samples protected from oxidation.

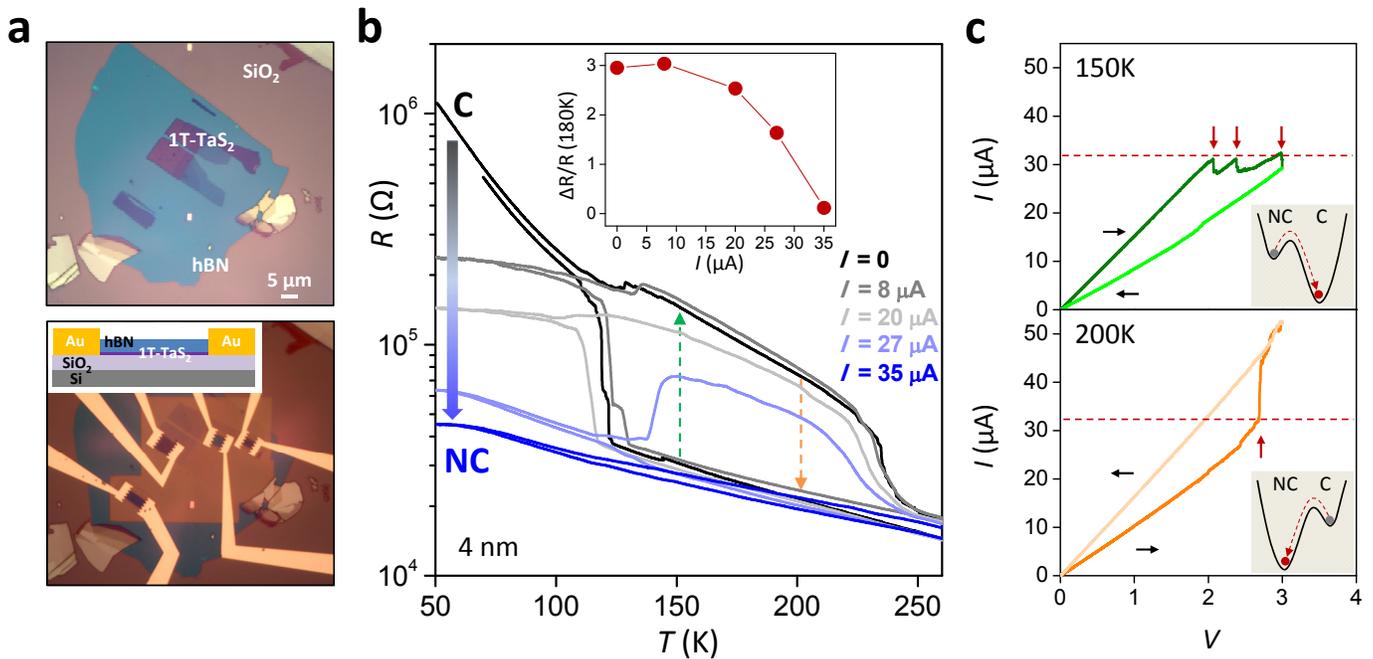

**Figure 2 | Electrical control of NC-C transition in oxidation-free, 2D devices. a.** Optical images of 1T-TaS$_2$ flakes on a SiO$_2$/Si wafer covered by hBN in inert atmosphere before (top) and after (bottom) side electrical contact. Inset shows side-view device schematic. **b.** AC resistance vs. temperature for 4nm thick device as a function of DC current. Continuous current flow stabilizes NC phase at low temperature. Normalized resistance difference between cooling and warming is plotted as a function of DC current in inset. **c.** (Top) Current vs. voltage sweep at 150K starting in NC phase shows abrupt decreases in current and transition to the C phase. (Bottom) Same at 200K starting in C phase shows abrupt increase in current and transition to NC phase. Free energy schematics of electrically induced transitions plotted in insets.

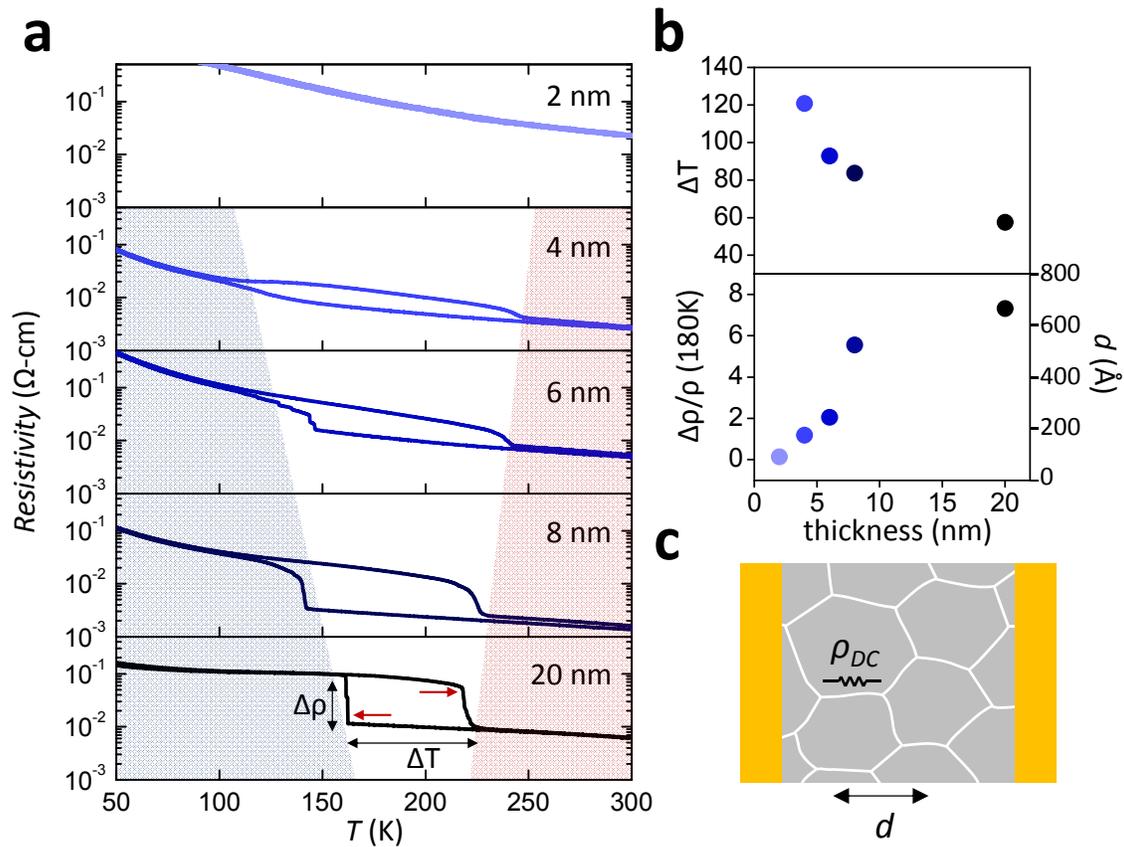

**Figure 3 | Dimensional dependence of phase transition (electron transport). a.** Thickness evolution of temperature-dependent resistivity around NC-C phase transition measured on hBN-covered ultrathin samples and 20nm thick flake. **b.** Temperature hysteresis (top) and normalized resistivity difference (bottom) between cooling and warming as a function of sample thickness. Hysteresis widens and resistivity difference decreases in thinner samples. Resistivity change can be used to estimate the discommensuration density $1/d$ at low temperature. **c.** Circuit model of discommensuration network.

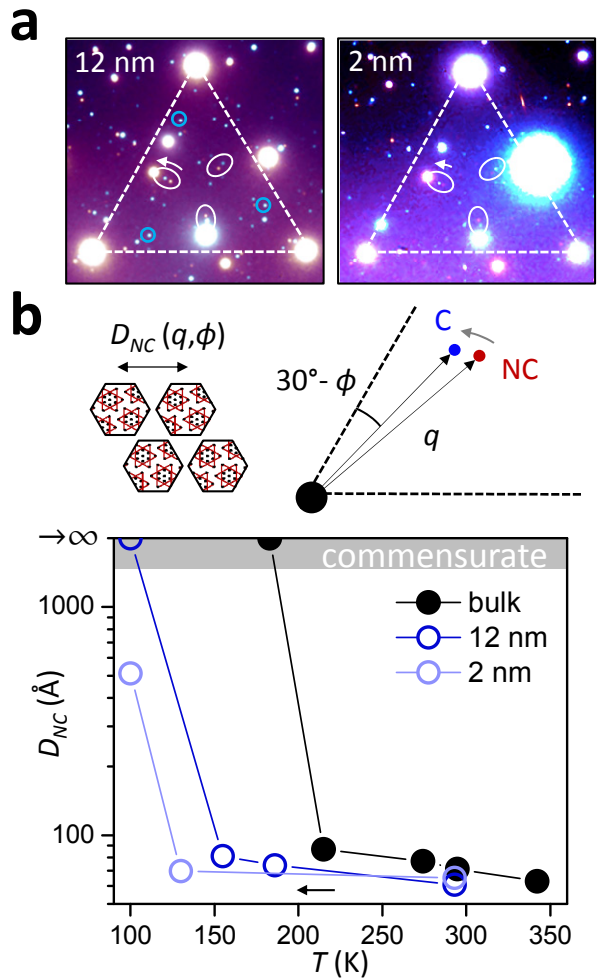

**Figure 4 | Dimensional dependence of phase transition (electron diffraction). a.** Overlaid TEM diffraction images of ultrathin 1T-TaS$_2$ covered with hBN taken at 295K (red peaks) and 100K (blue peaks) for two flake thicknesses. hBN preserves CDW order (circled peaks) but introduces additional diffraction spots. **b.** (Top right) Zoom-in schematic of CDW diffraction peaks showing temperature evolution. Position of NC spot can be used to estimate commensurate domain periodicity $D_{NC}$ (top left). (Bottom) $D_{NC}$ vs. temperature with cooling measured for the two covered samples compared with bulk data reproduced from ref. 13. Reduced thickness pushes NC to C phase transition to lower temperature.

# SUPPLEMENTARY INFORMATION

**Amorphous oxide surface layers:**

High-resolution, high-angle annular dark-field (HAADF-STEM) of the ultrathin 1T-TaS$_2$ sheets revealed the presence of amorphous surface layers. A curled sheet provided a cross-sectional view of the structure and clear evidence that the amorphous layers reside on the top and bottom surfaces (Fig. S1a). The sheet curls upward (Fig. S1b) such that the bottom amorphous layer does not overlap the 1T-TaS$_2$ sheets when viewed in projection, providing an ideal view of the amorphous layer. The layer is thicker than the 1T-TaS$_2$ interlayer spacing and has a lower atomic weight density as indicated by the layer's lower intensity in the HAADF-STEM image (a technique sensitive to atomic number and density). The amorphous layers are also visible in planar projection images (Fig. S1c), with a slow varying intensity (highlighted in Fig. S1d) that is clearly distinguishable from the 1T-TaS$_2$ periodic lattice.

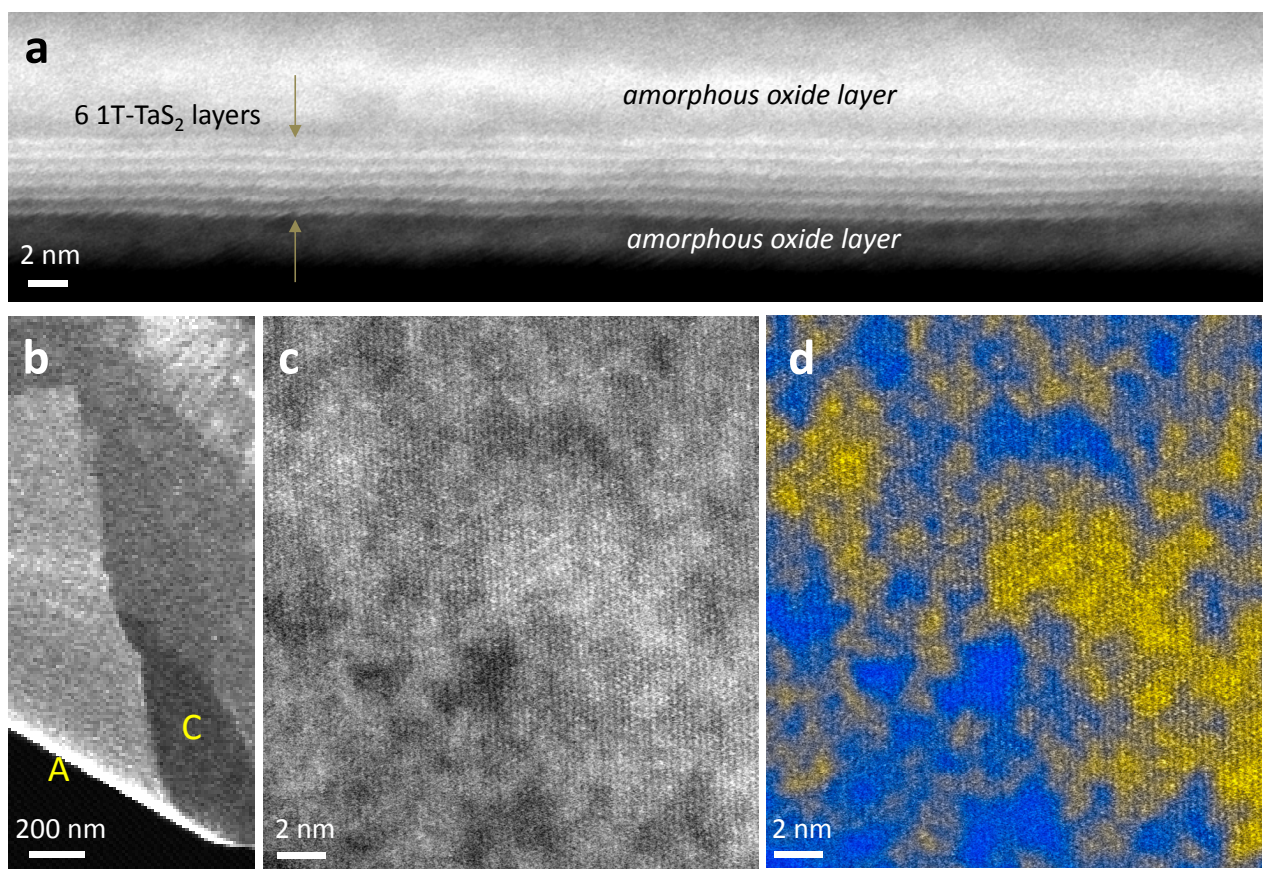

**Figure S1 | High-resolution STEM image of ultrathin (6 layers) 1T-TaS$_2$ sheet prepared in air. a.** Amorphous layers appear on the top and bottom surfaces. **b.** Image shows an overview of the curled sheet providing planar (c) and in-plane (a) viewing. **c.** High resolution STEM image of the 1T-TaS$_2$ sheet shows the high-frequency atomic structure and a lower-frequency intensity variation corresponding to the amorphous surface layers. The amorphous surfaces are more clearly visualized in **d.** which uses Lab Color space to create blue/yellow contrast of the amorphous (low-frequency) intensity variation.

Chemical analysis by both dispersive x-ray (Fig. S2a) and electron energy loss spectroscopy (Fig. S2b) taken from a planar projection confirms the presence of Ta, S, O and trace amounts of C. Tantalum is known to oxidize, with its most stable state as tantalum pentoxide (TaO$_5$). The oxygen signal is relatively strong and most likely indicates that the surface layers are an amorphous oxide. The presence of a carbon signal (large inelastic cross section) is weak relative to the oxygen signal (smaller inelastic cross-section) and may result from contamination—introduced during imaging within the electron microscope or existing previously.

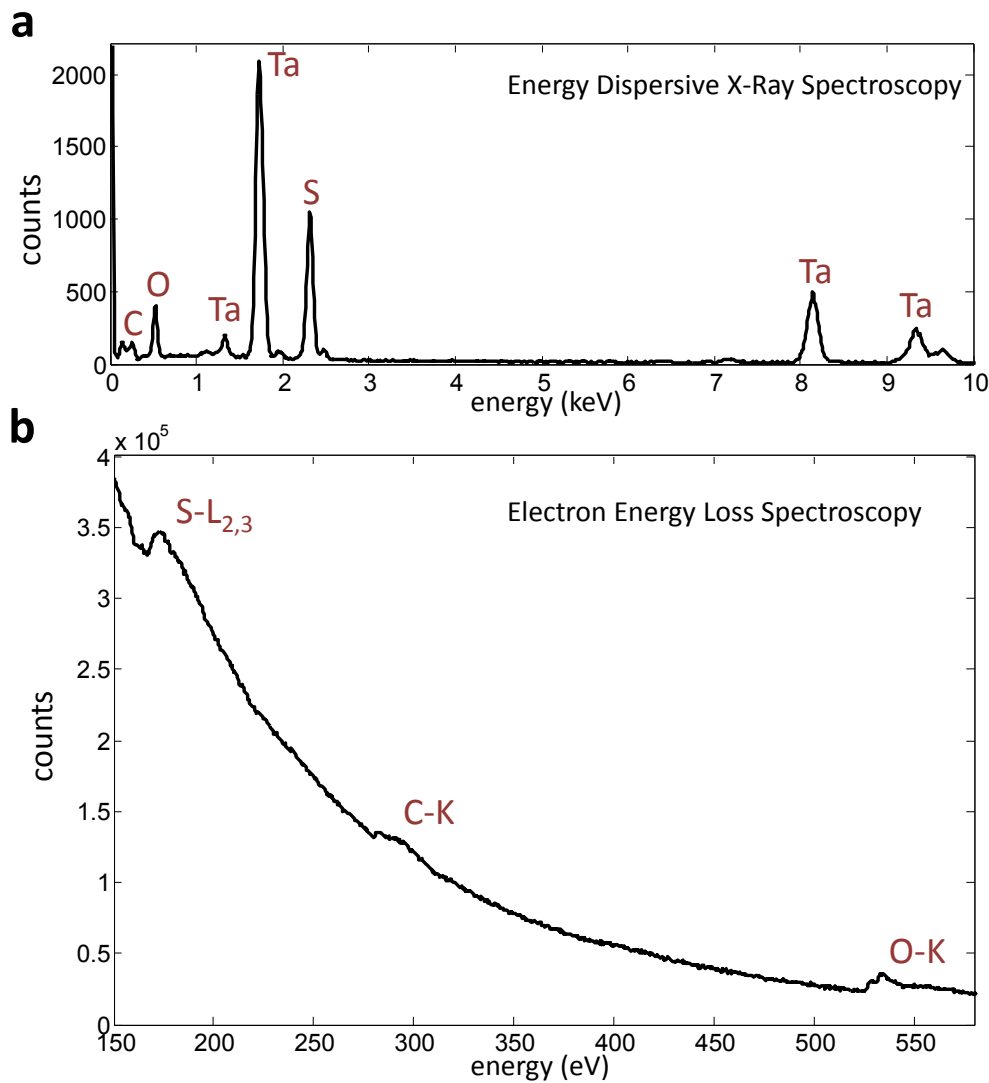

**Figure S2 | Chemical analysis with STEM-spectroscopy of ultrathin 1T-TaS$_2$ exfoliated in air.** In addition to the expected presence of Ta and S, oxygen and trace amounts of carbon are present in both the **a.** dispersive x-ray and **b.** electron energy loss spectroscopy. The sample was suspended such that all detected elements represent chemical species present in the specimen.

**Energy analysis of discommensuration pinning:**

We discuss the thickness dependence of the pinning energy of a discommensuration (DC) based on the arguments given in refs 1 & 2. The charge density wave (CDW) is described by a complex order parameter $\psi = |\psi|e^{i\theta}$. For simplicity, we assume a constant amplitude in the nearly commensurate (NC) phase. The energy is given by[3,4]

$$E = \frac{J}{2}(\nabla\theta - \delta q)^2 + W(1 - \cos M\theta),$$

where $\delta q$ is the deviation of the NC CDW wave vector from the commensurate value, and $W$ is the commensurate potential ($M$: integer). The first (elastic) term tries to match the phase of the CDW to the incommensurate value, $\theta = \delta qx$. On the other hand, the commensurate potential tends to pin the phase to an integer multiple of $\frac{2\pi}{M}$. As a result of these competing effects, the ground state, which minimizes the energy, is a periodic array of DCs between which the phase take the commensurate value—a DC is a local sudden change of phase from $\frac{2\pi p}{M}$ to $\frac{2\pi(p+1)}{M}$ ($p$: integer). The density of DCs is such that the phase propagates as $qx$ on a large scale, i.e. $n = \frac{qM}{2\pi}$. Since the modulation of charge density $\rho$ is related to the derivative of the CDW phase $\theta$ as

$$\rho = \frac{1}{\pi}\nabla\theta,$$

each DC carries an electric charge of $e^* = \frac{2e}{M}$. Due to this charge $e^*$, DCs tend to be pinned at impurities (for attractive interactions).

In a bulk sample, the DC becomes an extended planar object. We now estimate the pinning energy of a single DC plane under the influence of many impurities. The total energy is given by[1]

$$E = \frac{1}{2}K\int d\vec{r}\,(\vec{\nabla}z_0)^2 - V\sum_a \Theta(z_0(\vec{r}_a) - z_a),$$

where $K$ is the elastic constant, $z_0(\vec{r})$ is the location of a DC plane specified at point $\vec{r}$ in the $xy$-plane (see Fig. S3a). The second term represents the attractive interaction between a DC and impurities at $(\vec{r}_a, z_a)$. $\Theta$ is 1 when $|z_0(\vec{r}_a) - z_a| < \xi$ ($\xi$: the DC width), and 0 otherwise.

The random impurity potential tends to roughen the surface of the DC plane, and the fluctuation of an interface is characterized by two length scales, $L$ and $w$. The interface is nearly constant over $L$ along the interface, while it deviates by $w$ along the transverse direction. These are determined by minimizing the total energy. An important length scale is the mean impurity distance inside the DC (see upper panel, Fig. S3b),

$$l = 1/\sqrt{\xi n_{imp}},$$

where $n_{imp}$ is the three-dimensional impurity density. We consider the case where the thickness of the sample $t$ (along the $y$-axis) is shorter than $l$ (see lower panel, Fig. S3b). In this case, the system is quasi-one-dimensional, and the mean impurity distance $l_{1D}$ along the $x$-axis is given by

$$l_{1D} = 1/(\xi t n_{imp}).$$

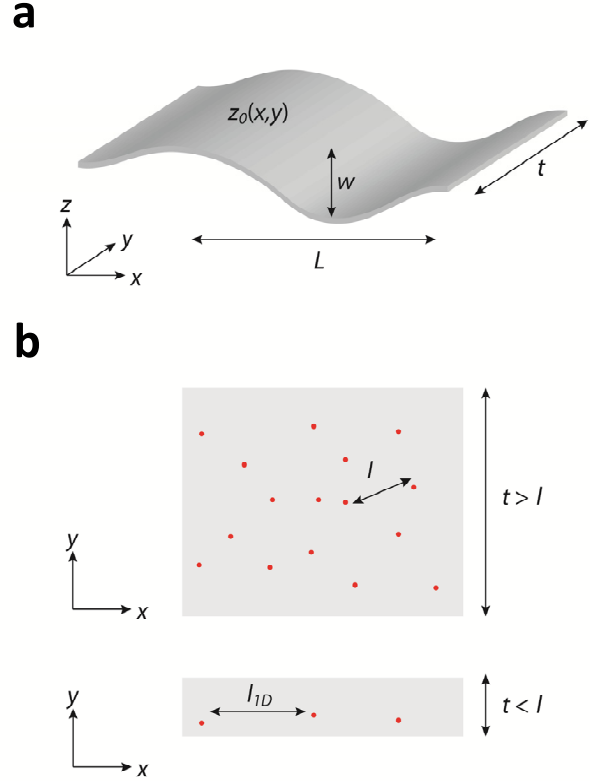

**Figure S3 | a.** Schematic picture of a DC plane and important length scales. In **b.**, red dots denote the location of impurities inside a DC plane. The effective mean impurity distance is $l$ for $t > l$, while it is $l_{1D}$ for $t < l$.

We now consider a DC plane in a domain of size $L \times t \times w$. We assume it contains many impurities along the $x$-direction ($L \gg l_{1D}$), and also that the DC is constant along $y$-direction. The elastic energy density is

$$\tfrac{1}{2} K \int_{-L/2}^{L/2} dx \int_0^t dy \; (\vec{\nabla} z_0)^2 (Lt)^{-1} \simeq \tfrac{1}{2} K w^2 L^{-2}.$$

The characteristic pinning energy is given by

$$E_{pin} = V \xi t L n_{imp}.$$

Considering that there are $w/\xi$ ways to put the DC in the domain, the maximum pinning energy is

$$-V \sqrt{\xi n_{imp} L^2 \ln\left[\tfrac{w}{\xi}\right]} (Lt)^{-1}.$$

Minimizing the total energy density in terms of $w$ and $L$, we find

$$w = \xi e,$$

$$L = \xi \left(\frac{4te^4K^2}{V^2 n_{imp}}\right)^{\frac{1}{3}}.$$

Now the pinning energy is

$$E_{pin} = -\frac{3}{4}\left(\frac{V^4 n_{imp}^2}{2e^2 t^2 K}\right)^{\frac{1}{3}} \propto t^{-\frac{2}{3}}.$$

And the number of impurity contained in the domain is

$$N = \xi^2 \left(\frac{4t^4 e^4 K^2 n_{imp}^2}{V^2}\right)^{\frac{1}{3}},$$

which is still large when $t \sim l$ in the weak-pinning limit, $K\xi^2 \gg V$. This justifies our assumption, $L \gg l_{1D}$. However, as $t \to 0$, $N$ becomes of the order of 1, which indicates that the DC is strongly pinned by impurities. This means that a bulk system in the weak pinning regime can be transformed into a strong pinning regime as the thickness becomes smaller than $l$.